\documentclass[a4paper,aps,twocolumn,english,floatfix,amsfonts,amssymb,superscriptaddress]{revtex4}
\usepackage{epstopdf}
\usepackage{graphicx}
\usepackage{dcolumn}
\usepackage{bm}
\usepackage{bbm,bbold}
\usepackage{color}
\usepackage{xcolor, soul}
\sethlcolor{green}
\usepackage{amsfonts}
\usepackage{amsmath}
\usepackage{mdframed}
\usepackage[normalem]{ulem}
\usepackage{mathrsfs}
\usepackage{lipsum, babel}
\usepackage[percent]{overpic}
\usepackage[colorlinks=true,citecolor=blue]{hyperref}
\hypersetup{colorlinks=true,citecolor=blue,linkcolor=blue,urlcolor=blue}
\usepackage{comment}
\let\oldAA\AA
\renewcommand{\AA}{\text{\normalfont\oldAA}}

\DeclareMathAlphabet\mathbfcal{OMS}{cmsy}{b}{n}
\begin{document}
\title{Nonlinear exciton drift in piezoelectric two-dimensional materials}

\author{Vanik Shahnazaryan}
\email{vanikshahnazaryan@gmail.com}
\affiliation{ITMO University, St. Petersburg 197101, Russia}

\author{Habib Rostami}
\email{habib.rostami@su.se}
\affiliation{Nordita, KTH Royal Institute of Technology and Stockholm University, Roslagstullsbacken 23, SE-106 91 Stockholm, Sweden}

\begin{abstract}
Noncentrosymmetric nature of single-layer transition metal dichalcogenides manifest itself in the finite piezoelectricity and valley-Zeeman coupling. We microscopically model nonlinear exciton transport in nano-bubble of single-layers of transition metal dichalcogenide. Thanks to the giant piezoelectric effect, we obtain an enormous internal electric field, $E_{\rm piezo}\sim 10^7$V/m, resulting in a built-in dipole moment of excitons. We demonstrate that the piezo-induced dipole-dipole interaction provides a novel channel for the nonlinear exciton transport distinct from the conventional isotropic funneling of excitons and leading to the formation of hexagon-shaped exciton droplet on top of a circularly symmetric nano-bubble. 
The effect is tunable via the bubble size dependence of the piezo-electric field $E_{\rm piezo} \sim h^2_{\rm max}/R^3$ with $h_{\rm max}$ and $R$ being the bubble height and radius, respectively. 
\end{abstract}

\maketitle

{\it Introduction.---} 
Single-layer (SL) of transition-metal dichalcogenides (TMDs) represent a flatland for probing rich exciton related phenomena \cite{ChernikovReview} owing to the direct band-gap in the visible frequency range.
Recently there is a rapidly growing interest towards exciton transport in SL-TMDs  \cite{Kulig2018}.
Various mechanisms governing the exciton transport were discussed, including the Seebeck effect \cite{Causin2019}, phonon drag \cite{Glazov2019} and spatially resolved Coulomb modulation of exciton energy \cite{Shahnazaryan2019}.
Exciton-exciton (XX) interaction (exciton nonlinearity), has strong impact on the exciton transport in conventional quantum wells \cite{Ivanov2002,Winbow2011,Cohen2011,Fedichkin2015,Dorow2016}. However, the many-body renormalization of  exciton transport in SL-TMD remains unexplored.
Exciton nonlinearity primarily manifests itself in the optical response through a blue shift in the exciton resonances \cite{Peyghambarian1984}. In unpolarized exciton gas the dominant interaction channel is  the short-range exchange \cite{Ciuti1998} which is also the case for pristine SL-TMDs \cite{Shahnazaryan2017,Barachati2018,Bleu2020,Stepanov2020,Shahnazaryan2020}. However, in case the excitons posses permanent dipole moment, the leading scattering channel  is the long-range dipole-dipole interaction \cite{Zhu1995,Berman2004,Kyriienko2012}.
The dipole moment can be induced in TMD excitons via an external electric field \cite{Pedersen2016,Engel2019,Chen2020} and in spatially indirect excitons in bilayers \cite{Gerber2019,Lorchat2020}. In this work, we discuss a novel mechanism for inducing exciton dipole moment in non-centrosymmetric SL-TMD due to strain.
\begin{figure}[t]
    \centering
    \includegraphics[width=0.99\linewidth]{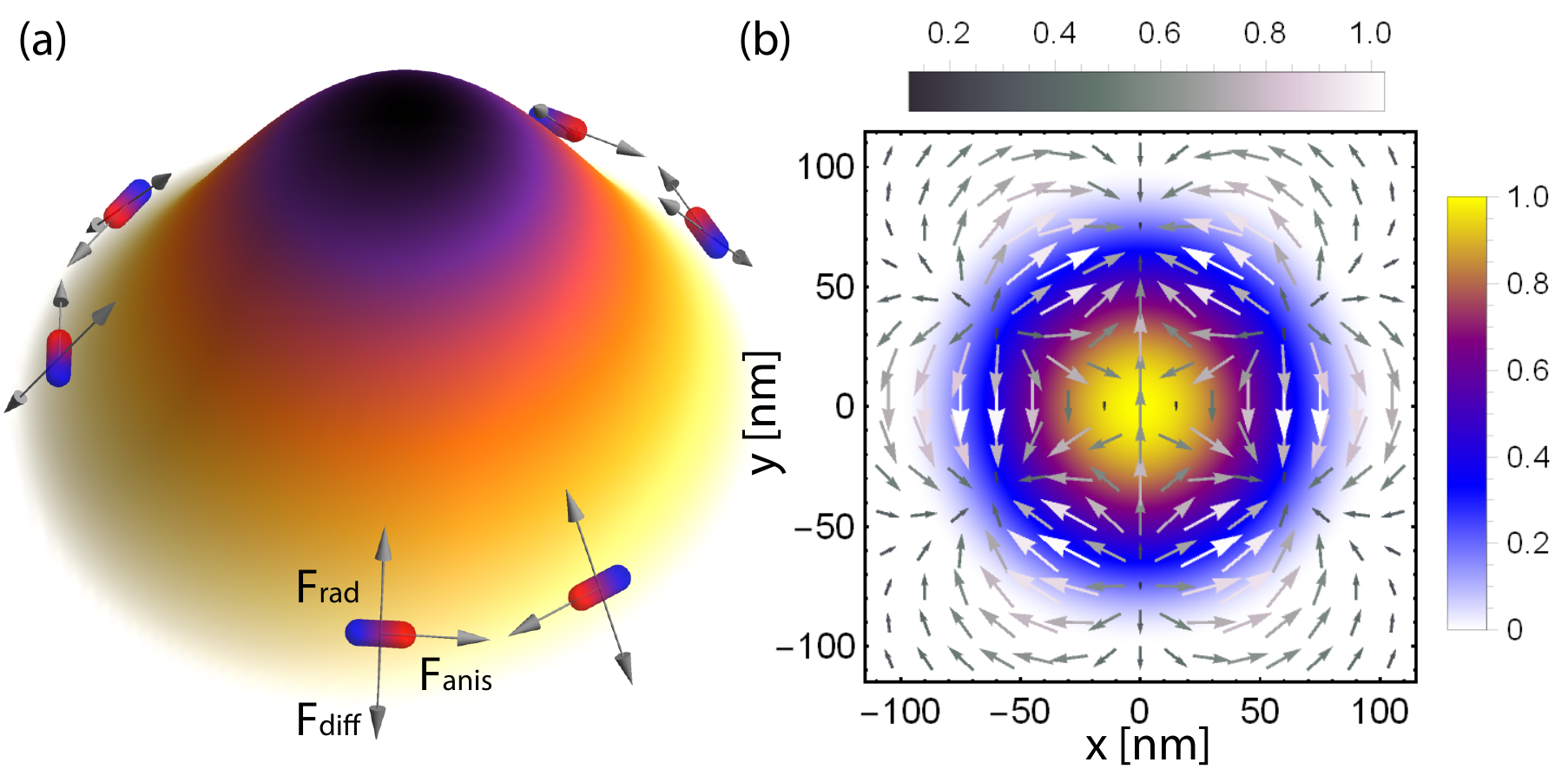}
    \caption{(a) The sketch of SL-TMD nano-bubble. The strain-induced  bandgap modulation gives rise to radial funneling force, which is partially compensated by the counteracting diffusive propagation. The piezoelectricity-induced dipolar interaction and the emergent pseudomagnetic field generate highly anisotropic forces, leading to spatially inhomogenous drift of excitons. (b) The electric field streamlines on top of bubble profile colormap. The horizontal (vertical) colorbar corresponds to piezo-induced electric field, $E_{\rm piezo}/E_0$ with $E_0 = 10^7$ V/m (height $h/h_{\rm max}$).}
    \label{fig:sketch}
\end{figure}
\begin{figure*}
    \centering
    \includegraphics[width=1.\linewidth]{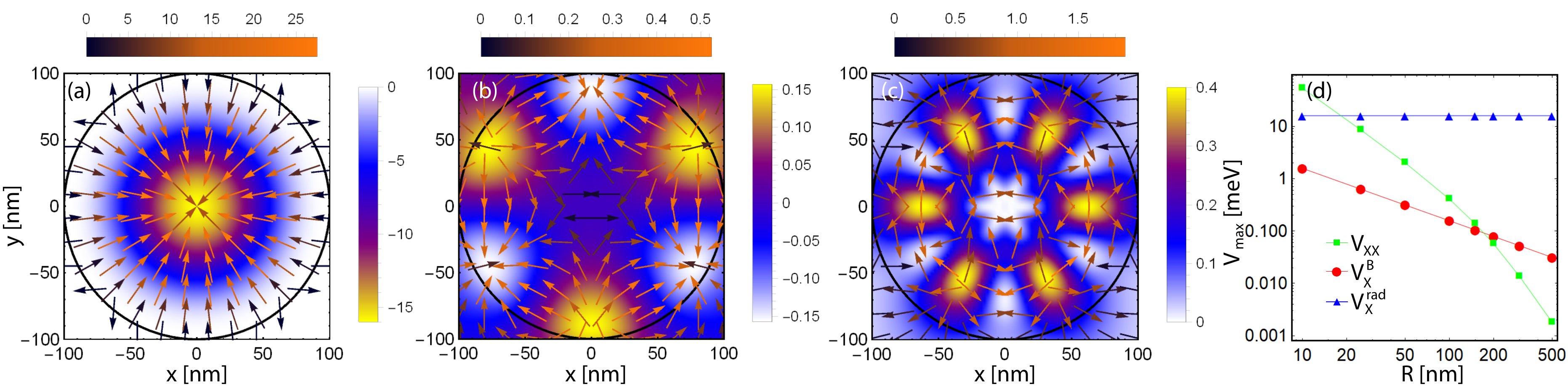}
    \caption{ 
    (a-c) The spatial colormap of strain-induced potential energies and the corresponding streamlines of drifting forces: (a) scalar funneling potential; (b) pseudo - magnetic field; (c) dipole-dipole interactions. The vertical (horizontal) colorbar corresponds to potential energy  in meV (drifting force in $F_0 = 100$ meV/nm).
    The black circle denotes the bubble boundary with radius $R =100$ nm.
    (d) Potential energy peak value plotted versus the bubble radius. For macroscopic bubbles ($R > 100$ nm) the exciton dynamics is solely governed by the radial funneling term while for nano-bubbles ($R < 100$ nm) the dipolar and magnetic sources rapidly increase by reducing the bubble radius.
    }
    \label{fig:pot}
\end{figure*}

Two-dimensional (2D) materials  are very flexible to out-of-plane deformation and strong to in-plane stretch \cite{Akinwande2017,Kim2019}. This unique property leads to nano-bubble formation in graphene and TMDs with a wide radius range 10 nm - 500 nm. The elastic stability of nano-bubbles enforces a universal aspect ratio of $h_{\rm max}/R\sim 0.1 - 0.2$ \cite{Khestanova2016,Blundo2020}, where $h_{\rm max}$ and $R$ stand for the height and radius of the bubble, respectively.
The non-uniform strain in nano-bubbles leads to spatial bandgap modulation \cite{Guo2020}, serving as an effective driving potential resulting in the exciton funneling effect \cite{Feng_nph_2012,Gomez2013}. The funneling manifests itself in a set of phenomena, such as the exciton nano-scale localization commensurate with exciton size \cite{Darlington2020}, strongly enhanced and localised photoluminescence \cite{Tyurnina2019}, and simultaneous direct and indirect bandgap photoluminescence \cite{Luo2020,Blundo2020}.

A manifestation of broken inversion symmetry in SL-TMDs is giant piezoelectric constant \cite{Wu2014,Zhu2015}.  Lattice deformation can displace the electronic Wannier centers from the background positive charge centers and therefore induce a non-vanishing polarization in the inversion-broken electrically insulating systems.
In SL-TMDs with hexagonal symmetry in the $xy$-plane, the piezoelectric polarization is given by ${\bf P}({\bf r}) = \gamma_{\rm piezo} \mathbfcal{A} ({\bf r}) \times \hat{\bf z}$  \cite{Droth2016,Rostami2018}, where ${\bf r}$ is the in-plane position coordinate, $\gamma_{\rm piezo}$ is the piezoeletic constant, and $\mathbfcal{A}$ is a fictitious gauge field given in terms of strain tensor component $({\cal A}_x,{\cal A}_y) =  (u_{xx}-u_{yy},-2 u_{xy})$  \cite{Rostami2018,Cazalilla2014} similar to case of graphene \cite{Guinea2010}.
The piezoelectric constant convey topological information about the valley-Chern number \cite{Rostami2018,Bistoni2019}.
For inhomogeneous strain profile there is also a pseudomagnetic field ${\bf B}^{(\tau)} =\partial_{\bf r}\times {\bf A}^{(\tau)}$ which changes sign in two valleys ($\tau=\pm$) at hexagonal Brillouin zone corner owing to the time reversal symmetry, ${\bf A}^{(\tau)}=\tau (\beta/2a_0)\mathbfcal{A}$ where $\beta$ is the Gruneisen parameter representing electron-lattice coupling and $a=\sqrt{3}a_0$ is the lattice constant, with $a=3.18$ \AA.
Another consequence of broken inversion symmetry in SL-TMDs is the valley-Zeeman effect \cite{Rostami2013,Li2014,MacNeill2015,Aivazian2015,Srivastava2015,Rostami2015} in the presence of an external magnetic field, ${\bf B} = B \hat {\bf z}$, where the exciton energy linearly modulates as $\tau g_{\rm VZ} \mu_B B$ where $\tau=\pm$ stands for the valley degree of freedom, $g_{\rm VZ}$ is the valley-Zeeman g-factor, and $\mu_B$ is the Bohr magneton.

In this Letter, we study nonlinear exciton transport in SL-TMDs driven by strong piezoelectric response. To the best of our knowledge, the interplay of XX-interaction, strong piezoelectricity and nano-bubble formation in TMD materials is not explored in the literature and our work aim to fill this gap. The considered system is schematically depicted in Fig.~\ref{fig:sketch} (a).
Nonuniform strain induces a bound charge density $\rho_{\rm piezo}({\bf r}) = - \partial_{\bf r} \cdot{\bf P} ({\bf r})$ that can generate a piezoeletric field based on the Poisson's equation:
\begin{align}
\label{eq:Epiezo}
\partial_{\bf r} \cdot \epsilon_0 {\bf E}_{\rm piezo}({\bf r}) = \rho_{\rm piezo}({\bf r})  \delta(z),
\end{align}
where $\epsilon_0$ is the vacuum permittivity.
Note that the Eq.~\eqref{eq:Epiezo} should be supplemented by Faraday's law $\partial_{\bf r}\times{\bf E}_{\rm piezo}=0$ enforcing the static character of the emergent electric field.
As discussed later, the giant piezoelectric effect yields in an enormous internal electric field of the order of $\sim 10^7$V/m, depicted in Fig~\ref{fig:sketch} (b), that can polarize excitons by inducing a dipole moment ${\bf d}({\bf r})= \alpha_E {\bf E}_{\rm piezo}({\bf r})$, where $\alpha_{E}$ stands for the exciton polarizability.
Utilizing the highly tunable piezo-induced dipole moment can be a novel framework for many-body driven exciton physics. Here, we microscopically develop a theory of exciton transport in strained SL-TMDs accounting for the long-range dipolar XX-interaction. Owing to the exciton nonlinearity, we predict a long-standing spatially three-fold symmetric exciton density which can be directly accessed via photoluminescence measurements.
  

%
\begin{figure*}
    \centering
    \includegraphics[width=1.\linewidth]{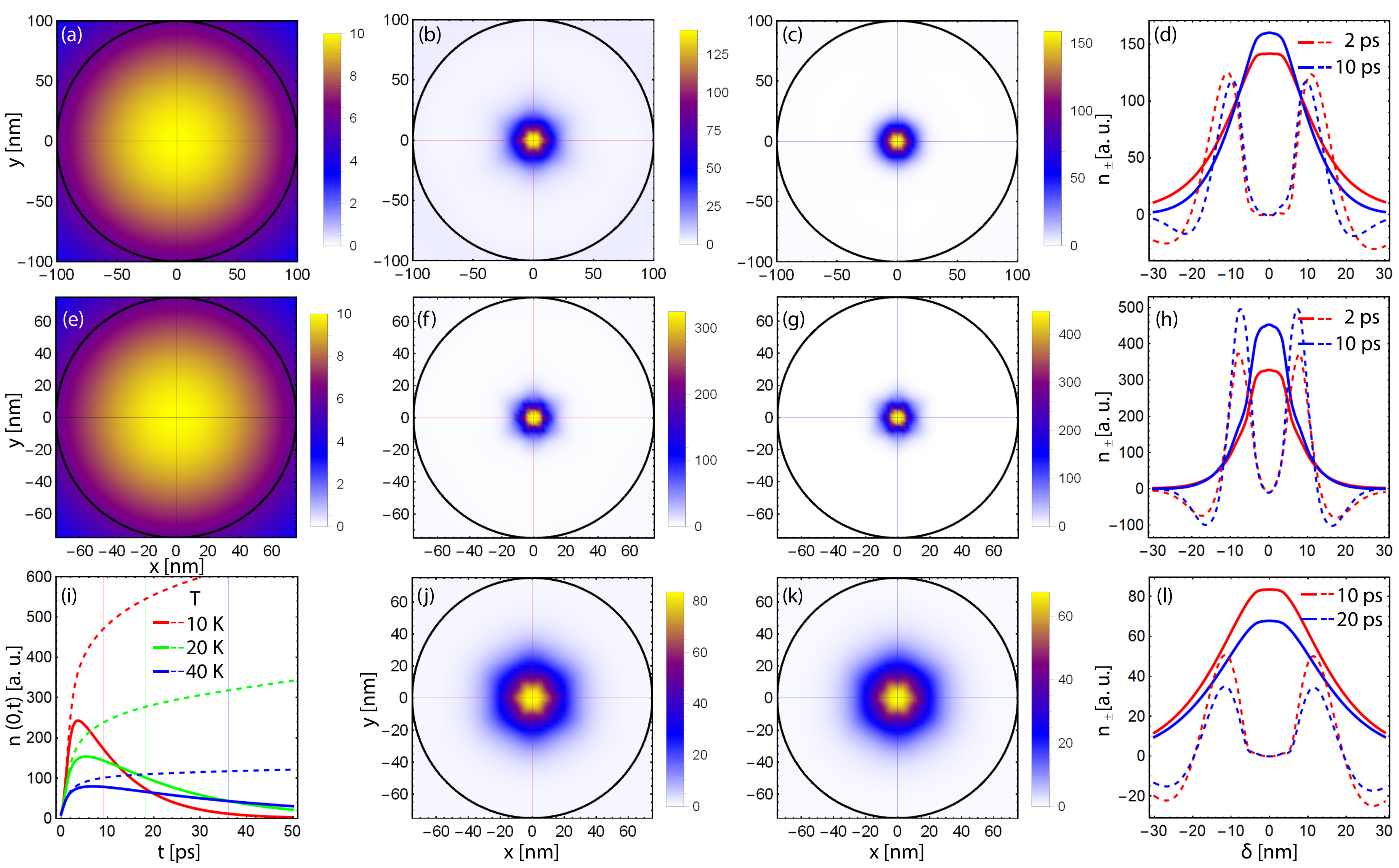}
    \caption{ (a), (b) and (c) snapshots of exciton density (in a.u.) for $T=10$ K and $R=100$ nm bubble radius at $t=$ 0, 2, 10 ps, respectively.
    The thin lines indicate cross-sections along $x$ and $y$ directions. 
    (d) The difference (dashed curves amplified by a factor of $10$) and the average (solid curve) of exciton density along $x$ and $y$ cross-sections at different time steps, corresponding to panels (b) and (c). 
    (e), (f) and (g) snapshots of exciton density (in a.u.) for $T=10$ K and $R=75$ nm bubble radius at $t=$ 0, 2, 10 ps, respectively. In smaller bubble the asymmetric hexagon-shaped is pronounced and persists for longer time. 
    (h) The difference (dashed curves amplified by a factor of $10$) and the average (solid curve) of exciton density in $x$ and $y$ cross-sections at different time steps, corresponding to panels (f) and (g).
    (i) exciton density at the bubble center in the absence of XX-interactions at several temperatures. The solid lines correspond to dissipative dynamics, and the dashed lines correspond to the absence of dissipation, i.e. $\tau_d\to \infty$. Vertical lines indicate the respective lifetime value $\tau_d=(0.1+0.9T[{\rm K}])$ps. 
    (j), (k) The snapshots of exciton density at $t=$ 10, 20 ps with elevated temperature $T=40$ K.
    (l) The difference (dashed curves amplified by a factor of $10$) and the average (solid curve) of exciton density in $x$ and $y$ cross-sections at different time steps, corresponding to panels (j) and (k).
   }
    \label{fig:dynamics}
\end{figure*}

{\it Microscopic theory of exciton transport.}--- 
The 2D exciton dynamics is characterized by Boltzmann distribution function $f_{\bf p}({\bf r},t)$, where {\bf p} is the exciton momentum. Accordingly, the exciton transport is modeled based on the well-known Vlasov-Boltzmann kinetic equation: 
\begin{align}
    &\partial_t f_{\bf p}({\bf r},t)  + {\bf v}_{\bf p}\cdot \partial_{\bf r} f_{\bf p}({\bf r},t)  
    +{\bf F}({\bf r},t)\cdot \partial_{\bf p} f_{\bf p}({\bf r},t) 
    \nonumber\\
   & =-\frac{f_{\bf p}({\bf r},t)}{\tau_d} +Q(f_{\bf p}), 
\end{align}
Notice that $\tau_d$, ${\bf v}_{\bf p}={\bf p} /M$ stand for the exciton lifetime and velocity, respectively, with $M$ being the exciton mass. The effective force ${\bf F}({\bf r},t)= -\partial_{\bf r} V({\bf r},t)$ in which $V({\bf r},t) =V_{\rm X}^{\rm rad}({\bf r}) +V_{\rm X}^{\rm B}({\bf r}) +V_{\rm XX}({\bf r},t)$ denotes the instantaneous potential energy acting on excitons at position ${\bf r}$.
The effective potential includes the exciton energy spatial modulation due to the strain-induced bandgap renormalization, $V_{\rm X}^{\rm rad}({\bf r})$; the valley-Zeeman shift due to the pseudomagnetic field, $V_{\rm X}^{\rm B}({\bf r})$; and the long-range dipole-dipole XX-interaction, $V_{\rm XX}({\bf r},t) = {\bf E}_{d}({\bf r},t) \cdot {\bf d}({\bf r})$. Here ${\bf E}_{d}({\bf r},t)$ is a mean-field electric field acting on excitons at  position ${\bf r}$ due to interaction with all other excitons reading as follows (see the Supplemental Material \cite{SM} for the details):
\begin{align}\label{eq:Ed}
\partial_{\bf r} \cdot \epsilon_0 {\bf E}_{d}({\bf r},t)= -\partial_{\bf r} \cdot [{\bf d}({\bf r}) n({\bf r},t) \delta(z)],  
\end{align}
where $n({\bf r},t)= \sum_{\bf p} f_{\bf p}({\bf r},t)$ stands for the exciton density. For a temporally slow varying electric field $\partial_{\bf r} \times {\bf E}_d({\bf r} ,t) \approx 0$, while for the exciton polarization density $\partial_{\bf r} \times [{\bf d}({\bf r}) n({\bf r},t) \delta(z)]$ does not necessarily vanish.  

In contrast to the real magnetic field, the pseudomagnetic field results in a normal Zeeman correction to the exciton energy owing to the absence of inversion and the presence time-reversal symmetry.  
Considering the conventional radial funneling potential owing to the bandgap renormalization that is proportional to the trace of strain tensor, we write the corresponding energy modulation as follows 
\begin{align}\label{eq:Vm_Vs}
V_{\rm X}^{\rm B}({\bf r}) = g_{\rm VZ} \mu_B B({\bf r}),~~~
~~~V_{\rm X}^{\rm rad}({\bf r}) = g_0 \sum_{\ell}u_{\ell\ell}({\bf r}).
\end{align}
We set the valley-Zeeman coupling $g_{\rm VZ}\approx -0.81$ \cite{Rostami2015} and the radial funneling strength  $g_{0}\approx 300$ meV \cite{Moon2020}.
We neglect the diamagnetic shift in \eqref{eq:Vm_Vs} as for the ground state exciton it is negligibly small compared with valley Zeeman effect \cite{Stier2018}.
The collision term $Q(f_{\bf p})$ which models the dephasing processes is treated within a phenomenological relaxation time approximation: $Q(f_{\bf p})=\left[\bar{f}_{\bf p}({\bf r},t) - f_{\bf p}({\bf r},t) \right]/\tau_C$ with $\tau_C$ denoting the collision (dephasing) time. Here $\bar{f}_{\bf p}({\bf r},t) \propto n({\bf r},t) e^{-E^{ kin}_{\bf p}/ (k_{\rm B} T)}$ 
follows the quasi-equilibrium Maxwell-Boltzmann distribution, with $E^{kin}_{\bf p}=p^2/ 2M$ as the center of mass kinetic energy of excitons, $k_{\rm B}$ denoting the Boltzmann constant and $T$ is the lattice temperature.
We utilize a Chapman–Enskog type ansatz \cite{Transport_Book2009} separating the equilibrium and non-equilibrium fractions of Boltzmann function  $f_{\bf p} = \bar f_{\bf p}+ g_{\bf p}$, with $\sum_{\bf p} g_{\bf p} =0$ 
and after the momentum integration we find \cite{SM} 
\begin{eqnarray} \label{eq:Drift-Diff}
\left [\partial_t +1/\tau_d \right] 
n({\bf r},t) =  \partial_{\bf r} \cdot \left[ D \partial_{\bf r} n({\bf r},t)  -\mu  n({\bf r},t)  {\bf F}({\bf r},t)  \right],~~
\end{eqnarray}
where $D=\mu k_{\rm B} T$ stands for the diffusion coefficient and $\mu= \tau_{\rm eff}/M$ is the mobility in which 
$1/\tau_{\rm eff}=1/\tau_C+1/\tau_d$ is an effective relaxation rate. Accordingly, the exciton mobility depends on collision time as well as the finite lifetime, while previously it its dependence on the lifetime was underestimated.  
For the short-range exchange interactions, i.e. $V({\bf r},t)= U_0 n({\bf r},t)$, the Eq. \eqref{eq:Drift-Diff} reduces to a form phenomenologically discussed in studies of exciton transport \cite{Fedichkin2015,Dorow2016}. 


{\it Characteristics of circular nano-bubble in TMDs.}---
Exact solution of displacement profile for  SL-TMD circular nano-bubble strongly depends on the elastic boundary conditions and external mechanical force distribution \cite{Rostami2018,Dai2018}, see also \cite{TimoshenkoBook,PitaevskiiBook}. 
However, using an intuitive perspective we can estimate a basic analytic solution. For instance, for the case of pure bending approximation, where the bending (curvature) energy dominates the elastic stretching energy \cite{Rostami2018}, the out-of-plane displacement is given in the form  
\begin{align}
h(r) = h_{\rm max}  \left(1-r^2/R^2 \right)^2 \Theta(R-r),
\end{align}
where $R$ and $h_{\rm max}$ are the radius and height of the bubble, which in our numeric we set the aspect ratio factor $\xi=h_{\rm max}/R=0.2$. To ensure the elastic stability there is also a radial displacement $u(r)$ to be determined. Considering circular symmetry of the bubble one can evaluate strain tensor elements in the polar coordinate as $u_{rr} =\partial_r u(r) + [\partial_r h(r)]^2/2$, $u_{\theta\theta}= u(r)/r$, and $u_{r\theta}=u_{\theta r}=0$. Utilizing linear elasticity formalism \cite{Rostami2018},  the radial displacement reads $u(r)= u_0  (4\alpha-7\alpha^7-20\alpha^5-18\alpha^3)$ for $\alpha\leq1$ and  $u(r)=-u_0/\alpha^2$ for $\alpha>1$  where $\alpha= r/R$ and $u_0= \xi h_{\rm max}/6$. The corresponding profile of bubble is shown in Fig.~\ref{fig:sketch} (b). Eventually, the trace of tensor strain reads 
$\sum_{\ell} u_{\ell\ell} =u_{rr}+u_{\theta\theta}$.
Having strain tensor components, we obtain pseudogauge vector $\mathbfcal{A}({\bf r})$ and thus the piezoelectric charge density follows 
\begin{align}
\rho_{\rm piezo}({\bf r})= - \partial_{\bf r} \cdot{\bf P}({\bf r}) =C\frac{\gamma_{\rm piezo}}{R} \rho\left(\frac{r}{R}\right) \sin(3\theta)
\end{align}
where $C=-4\xi^2/3$ and $\rho(\alpha)$ is a dimensionless function: 
$\rho(\alpha) =\alpha^3(4-3\alpha^2)$ for $\alpha<1$ and $\rho(\alpha) = 1/\alpha^3$ for $\alpha>1$. 
Plugging piezo-induced charge density into the Poisson equation given in Eq.~(\ref{eq:Epiezo}),  we evaluate the built-in electric field due to piezoelectricity. The resulting electric field is depicted in Fig.~\ref{fig:sketch} (b). Remarkably, it has a three-fold symmetry and is of order of $E_{\rm piezo}\sim 10^7$ V/m, in line with with recent experimental report of piezoelectricity in TMD nano-bubbles \cite{Palma2020}. 
Finally, it is easy to show that $\partial_{\bf r} \cdot (\mathbfcal{A}({\bf r})\times\hat{\bf z}) = \hat{\bf z}\cdot (\partial_{\bf r}\times\mathbfcal{A}({\bf r}))$ which implies $B({\bf r}) = \beta\rho_{\rm piezo}({\bf r})/(2a_0\gamma_{\rm piezo}) $. We set electron-lattice coupling to $\beta=3$. The rest of the paper is to numerically solve coupled nonlinear drift-diffusion equations (\ref{eq:Ed}), (\ref{eq:Vm_Vs}) and (\ref{eq:Drift-Diff}), thus we obtain dynamical density distribution and investigate the impact of different force sources.


{\it Anisotropic nonlinear exciton drift in TMD nano-bubble.}---
We numerically simulate the Eq.~\eqref{eq:Drift-Diff} to study nonlinear exciton transport in SL-TMDs. The parameters are chosen as collision time $\tau_C = 0.26$ ps \cite{Cadiz2018}, piezelectric constant $\gamma_{\rm piezo} = 2.9 \times 10^{-10} $ C/m \cite{Wu2014,Zhu2015}, and exciton polarizability  $\alpha_E = 5  \times 10^{-18}$ eV(m/V)$^2$ \cite{Pedersen2016}. The exciton lifetime scales linearly with temperature, i.e. $\tau_d \propto T$ \cite{Andreani1991}, and for SL MoS$_2$ it can be approximated as $\tau_d \approx (0.1 + 0.9 T {\rm [K]} ) $ ps \cite{Palummo2015}. We set the radius of bubble as $R= 100$ nm, and assume an initial exciton population density as $n({\bf r},0) = n_{\rm max} e^{-|{\bf r}-{\bf r}_0|^2/\Delta^2}$, where we set $\Delta= R$, $n_{\rm max} = 10^{13}$ cm$^{-2}$ \cite{Moody2016} and with the center of pump spot coinciding with the bubble center, i.e. ${\bf r}_0 = 0$. 

The Fig.~\ref{fig:pot} demonstrates the spatial landscape of drifting potentials illustrating the conventional radial funneling, the trigonal symmetric Zeeman coupling and XX-interaction, respectively in panels (a), (b) and (c). The XX-interaction potential originates from the asymmetric profile of piezoelectric induced electric field depicted in Fig.~\ref{fig:sketch} (b). The corresponding force vector lines are depicted on each colormap plot. 
Remarkably, the XX-interaction force is significant close to the center of the bubble, unlike the negligibly small magnetic-field induced force. 
Considering the universal aspect ratio factor $\xi$, the only control parameter is the bubble radius $R$. Accordingly, we plot the radius dependence of each drifting potential in Fig.~\ref{fig:pot} (d) where it depicts the maximum value of each term.
As seen, the drifting potential scales with the bubble radius as  $1/R^\eta$ with $\eta=0,1$ and $\sim 2$ for $V^{\rm rad }_{\rm X}$, $V^{\rm B }_{\rm X}$ and $V_{\rm XX}$, respectively. For large size bubbles, e.g. $R>100$ nm, the radial funneling term is the dominant driving force compared  to the magnetic and  XX-interaction terms. Based on the different scaling of potentials with $R$, in nano-bubbles, e.g. $R<100$ nm, the magnetic and  XX interactions induced force are enhanced to compete with the radial funnel term, leading to an anisotropic exciton density.

The real time of exciton transport is presented in Fig.~\ref{fig:dynamics} where panels (a), (b), and (c) indicate snapshots of exciton distribution at $T=10$ K corresponding to $t=$0, 2 ps, and 10 ps, respectively. As seen, at the intermediate stage the particle distribution is strongly asymmetric showing a hexagonal profile [see panel (b)].
At the later time, panel (c), this asymmetry becomes less pronounced due to the finite exciton lifetime. 
Cross-section plots along $x$ and $y$-directions are depicted in panel (d) which illustrate the asymmetry of exciton density on a circularly symmetric bubble more quantitatively. 
Note that we adapt notation $n_{\pm}(\delta,t) = [n(x=\delta,y=0,t) \pm n(x=0,y=\delta,t)]/2$ for the population average ($+$) and difference ($-$) with corresponding solid and dashed lines, respectively. The anisotropic density profile is revealed by a vanishing $n_{-}$ at the bubble center and two sharp peaks at the edges of the central hexagon. The distance between two peaks $2\delta_0$ can be a good experimental probe to estimate the size of the hexagon side $\approx \delta_0$. 
Due to a fast initial radial drift of particles to the bubble center, the pseudomagnetic field negligible in the bubble center has vanishing impact on the particle distribution.
Hence, the anisotropic exciton density distribution at the bubble central region is only associated with XX-interaction effect.

To preserve the asymmetric density distribution for a longer time, we decrease bubble radius to $R =75$ nm. The snapshots of evolution are shown in Fig.~\ref{fig:dynamics} (e)-(g). One can observe the hexagon-like distribution last for the longer period by tracing the similarity of density profile at $t=2$ ps and $t=10$ ps shown in panel (f) and (g), respectively. The long-living asymmetry is clearly visible in looking at the time-evolution of population difference along cross-sections, c.f. the panels (h) and (d). The pronounced asymmetry of density distribution is due to the enhancement of XX-interaction for the smaller bubble size. The inhomogeneous distribution of excitons can be retained even longer in time by increasing the temperature. First of all, raising temperature $T$ will increase the exciton lifetime. Secondly, it controls the diffusion process where at higher temperature the diffusion constant will be enhanced. The counter oriented diffusion effect (from center to bubble edge) can partially compensate the radial funnel effect (from edge to the bubble center) by altering the temperature.
To unveil the impact of diffusion and such a partial compensation, we analyze the dynamics of excitons in the absence of anisotropic forces at different temperatures after ignoring dipole-dipole and magnetic potential energies. The resulting exciton population in the bubble center is shown versus time in Fig.~\ref{fig:dynamics} (i). Remarkably, in the case when the dissipation is neglected (dashed curves) a temperature dependent dramatic drop in the exciton density temporal slope is predicted. Moreover, at elevated temperatures such a quasi-equilibrium dynamics is reached at timescale far below the exciton lifetime [the blue dashed curve]. The corresponding evolution at $T=40$ K is shown in Fig.~\ref{fig:dynamics} (j) - (l).
Given by the partial compensation of radial funneling by the diffusion effect, a nearly constant hexagonal shape lasts during the evolution process. After including all driving forces, the density snapshots at $t=10$ ps and $t=20$ ps for bubble size $R=75$ nm and temperature $T=40$ K are depicted in panels (j) and (k), respectively. The corresponding cross-section plot is shown in panel (l). As seen the net effect of raising temperature is an increase in the central hexagon size with sharper edges which can last for a longer time.


{\it Summary and Outlook.}--- The piezoelectricity impact on the nonlinear exciton transport in SL-TMD nano-bubble is studied. We develop a nonlinear drift-diffusion theory for excitons starting from the Vlasov-Boltzmann kinetic equation.
Strong piezo-induced electric filed can polarize excitons 
resulting in a novel interaction channel being available in the nano-bubble systems. Furthermore, we discuss other drifting forces originating from strain-induced bandgap renormalization and the valley-Zeeman coupling.
In nano-bubbles of SL-TMD the piezo-induced XX-interaction is the dominant factor in describing the long-living hexagon-shaped of the exciton droplet. The hexagon size and time evolution are controllable via the temperature dependence of diffusion and the bubble radius dependence of the XX-interaction. Our technical theory can be generalised in other systems such as hetero-structures of 2D materials with different inhomogeneous strain profiles or exhibiting Moiré patterns \cite{Yuan2020}. Based on the current nano-fabrication technology a lattice of long-range interacting droplets in an array nano-bubbles in TMD is designable, potentially   resulting in a spontaneous long-range coherence.  

\section*{Acknowledgments}
H.R. thanks Emmanuele Cappelluti for useful discussions.
This work was supported by the Russian Science Foundation (grant No. 19-72-00171).
H.R. acknowledges the support from the Swedish Research Council (VR 2018-04252).

\newpage
\begin{widetext}


\section*{Supplemental Material}

\subsection{Mean-field electric field induced by a dipolar exciton gas}

The charge density of a single  dipole reads as
\begin{align}
    \rho_j ({\bf r}) &= \lim_{ \Delta{\bf r}_j \to  {\bf 0}} q \left [ \delta({\bf r}-{\bf r}_j - \Delta{\bf  r}_j ) -\delta({\bf r}-{\bf r}_j ) \right ] \delta(z)  \notag \\
    &= -\lim_{\Delta {\bf r}_j \to {\bf 0}} (q \Delta {\bf r}_j) \cdot \partial_{{\bf r}}  \delta({\bf r}-{\bf r}_j )  \delta(z) \notag \\
    & \equiv- {\bf d}({\bf r}_j) \cdot \partial_{{\bf r}}  \delta({\bf r}-{\bf r}_j )  \delta(z).
\end{align}
Notice that ${\bf d}({\bf r}_j) = \lim_{\Delta {\bf r}_j \to {\bf 0}} (q \Delta {\bf r}_j)$ is the dipole moment located at position ${\bf r}_j$ and $q$ is a unit of electric charge. Therefore, the electric field created by a single dipole satisfies the following Poisson's equation
\begin{align}
    \partial_{{\bf r}} \cdot \epsilon_0 {\bf E}({\bf r},{\bf r}_j)=\rho_j ({\bf r}) =  - {\bf d}({\bf r}_j)  \cdot \partial_{{\bf r}}  \delta({\bf r}-{\bf r}_j )  \delta(z).
\end{align}
The total electric field created by an ensemble of dipoles is
\begin{align}
    {\bf E}_d ({\bf r} ) = \sum_{j}{\bf E} ({\bf r},{\bf r}_j)  = \int {\bf E}({\bf r},{\bf r}' ) n({\bf r}') {\rm d} {\bf r}' ,
\end{align}
where we used the definition of exciton density $n({\bf r}) = \sum_{j} \delta ({\bf r} -{\bf r}_j )$ with $\sum_j$ summing over all excitons. Thus, for the total electric field we reach 
\begin{align}
    \partial_{{\bf r}} \cdot \epsilon_0 {\bf E}_d ({\bf r})=\sum_j \rho_j ({\bf r}) &= -
    \int n({\bf r}') {\bf d} ({\bf r}')  \cdot  \partial_{{\bf r}}  \delta({\bf r}-{\bf r}' )  \delta(z) {\rm d} {\bf r}' 
    \notag \\
   & = -
    \partial_{{\bf r}}   \cdot  \int n({\bf r}') {\bf d} ({\bf r}')   \delta({\bf r}-{\bf r}' )  \delta(z) {\rm d} {\bf r}' 
       \notag \\
    & = -\partial_{{\bf r}} \cdot \left[ {\bf d}({\bf r}) n({\bf r}) \delta(z) \right] ,
\end{align}
By replacing the static density $n({\bf r})$ with the instantaneous time-dependent density $n({\bf r},t)$ we obtain 
\begin{align}
    \partial_{{\bf r}}\cdot  \epsilon_0 {\bf E}_d ({\bf r},t) 
     = -\partial_{{\bf r}} \cdot \left[ {\bf d}({\bf r}) n({\bf r},t) \delta(z) \right] ,
\end{align}
corresponding to Eq.~\eqref{eq:Ed} of the main text.

\subsection{From Vlasov-Boltzmann kinetic theory to nonlinear drift-diffusion equation}

In order to derive the drift-diffusion equations, we proceed with dimensionless variables. To do so, we introduce characteristic quantities describing the system. The characteristic velocity is determined by the lattice temperature as $v_0=\sqrt{k_B T/M}$. Hence, the mean free path reads as $\lambda = v_0 \tau_C$. The time, which a particle with the typical velocity $v_0$ needs to run through the sample, is $\tau_0 =L/v_0$, where $L$ is the length of the sample. One can characterize the system with the reference length $\lambda_0= \sqrt{\lambda L}$. 
Finally, the reference momentum is $p_0 =M v_0 =\sqrt{M k_B T}$. We introduce dimensionless quantities as $    \Tilde{{\bf r}}={\bf r} /\lambda_0$, $\Tilde{V}=V/(k_B T)$, $\Tilde{{\bf p} }={\bf p}/p_0$, $\Tilde{\bf v}_{\bf p}= {\bf v}_{\bf p}/v_0 = \Tilde{{\bf p}}$, $\sigma=\lambda/\lambda_0$, $\Tilde{t}=t/\tau_0= t \sigma^2 /\tau_C $, $\Tilde{Q}(f)=\tau_C Q(f)$.
Then in dimensionless form the kinetic equation will read as 
\begin{align}
    \sigma^2 \partial_{\Tilde{t}}  f_{\bf \Tilde{p} } (\Tilde{\bf r},\Tilde{t}) 
    +\sigma \left[  \Tilde{\bf p} \cdot \partial_{\Tilde{\bf r}} f_{\Tilde{\bf p}}  
    -  \partial_{\Tilde{\bf r}} \Tilde{V} (\Tilde{\bf r},\Tilde{t}) \cdot  \partial_{\Tilde{\bf p}} f_{\Tilde{\bf p}} (\Tilde{\bf r},\Tilde{t}) \right] +\frac{\tau_C}{\tau_d} f_{\Tilde{\bf p}} (\Tilde{\bf r},\Tilde{t})
    =\Tilde{Q}(f_{\Tilde{\bf p}}(\Tilde{\bf r},\Tilde{t})) . 
\end{align}
We apply a Chapman–Enskog type ansatz \cite{Transport_Book2009} 
as $f_{\Tilde{\bf p}} (\Tilde{\bf r},\Tilde{t}) = \left[ n_0(\Tilde{\bf r},\Tilde{t}) \bar{f}_{\Tilde{\bf p}}^0 +\sigma g_{\sigma\Tilde{\bf p}} (\Tilde{\bf r}, \Tilde{t}) \right] e^{-\frac{\tau_0}{\tau_d} \Tilde{t}}$, which account for the finite lifetime of excitons. 
Here $\bar{f}_{\Tilde{\bf p}}^0=2\pi e^{-\frac{|\Tilde{p}^2|}{2}}$, and $g_{\sigma\Tilde{\bf p}}$ is a small correction to homogeneous distribution. 
Inserting into kinetic equation, we get
\begin{align}
    & \sigma \left[ \left(\partial_{\Tilde{t}} n_0 (\Tilde{\bf r},\Tilde{t}) - \frac{\tau_0}{\tau_d} n_0 (\Tilde{\bf r},\Tilde{t})  \right) \bar{f}_{\Tilde{\bf p}}^0      + \sigma \left( \partial_{\Tilde{t}} g_{\sigma \Tilde{\bf p}} (\Tilde{\bf r},\Tilde{t})  - \frac{\tau_0}{\tau_d} g_{\sigma \Tilde{\bf p}} (\Tilde{\bf r},\Tilde{t}) \right)  \right] +
    \Tilde{\bf p} \cdot \partial_{\Tilde{\bf r}} n_0  (\Tilde{\bf r},\Tilde{t}) \bar{f}_{\Tilde{\bf p}}^0  
    - \partial_{\Tilde{\bf r}} \Tilde{V} (\Tilde{\bf r},\Tilde{t}) \cdot  \partial_{\Tilde{\bf p}} n_0 (\Tilde{\bf r},\Tilde{t}) \bar{f}_{\Tilde{\bf p}}^0 
    \notag \\
    & +\sigma \left[  \Tilde{\bf p} \cdot \partial_{\Tilde{\bf r}} g_{\sigma\Tilde{\bf p}} (\Tilde{\bf r},\Tilde{t})
    - \partial_{\Tilde{\bf r}} \Tilde{V} (\Tilde{\bf r},\Tilde{t}) \cdot  \partial_{\Tilde{\bf p}} g_{\sigma\Tilde{\bf p}} (\Tilde{\bf r},\Tilde{t}) \right] 
    +\frac{\tau_C}{\tau_d \sigma} n_0 (\Tilde{\bf r},\Tilde{t}) \bar{f}_{\Tilde{\bf p}}^0 
    + \frac{\tau_C}{\tau_d} g_{\sigma\Tilde{\bf p}} (\Tilde{\bf r},\Tilde{t}) =-g_{\sigma\Tilde{\bf p}} (\Tilde{\bf r},\Tilde{t}),
    \label{eq:Chapman}
\end{align}
where we use $\Tilde{Q}(g_{\sigma\Tilde{\bf p}}) =-\sigma g_{\sigma\Tilde{\bf p}} (\Tilde{\bf r},\Tilde{t})$.
Here we recall that $\sigma=\lambda/\lambda_0=\sqrt{\tau_C/\tau_0}$, leading to
$\sigma \tau_0 /\tau_d = \tau_C/(\sigma \tau_d)$, resulting in the cancellation of corresponding terms. 
In the limit $\sigma \ll 1$ one has
\begin{align}
    g_{\sigma \Tilde{\bf p}} (\Tilde{\bf r},\Tilde{t}) = -\frac{1}{1+\tau_C/\tau_d} 
    \left[\Tilde{\bf p} \cdot \partial_{\Tilde{\bf r}} n_0 (\Tilde{\bf r},\Tilde{t}) \bar{f}_{\Tilde{\bf p}}^0  
    -  \partial_{\Tilde{\bf r}} \Tilde{V}  (\Tilde{\bf r},\Tilde{t}) \cdot  \partial_{\Tilde{\bf p}} n_0 (\Tilde{\bf r},\Tilde{t}) \bar{f}_{\Tilde{\bf p}}^0 \right] . 
\end{align}
We note that $g_{\sigma \Tilde{\bf p}}$ is an odd function of momentum, and doesn't contribute to the density distribution.
We further integrate over momentum the Eq. \eqref{eq:Chapman}  
and note that the terms 
$\langle \Tilde{\bf p} \cdot \partial_{\Tilde{\bf r}}  n_0 (\Tilde{\bf r},\Tilde{t})  \bar{f}_{\Tilde{\bf p}}^0  \rangle$, 
$\langle \partial_{\Tilde{\bf p}}  n_0 (\Tilde{\bf r},\Tilde{t}) \bar{f}_{\Tilde{\bf p}}^0  \rangle $, $\langle g_{\sigma \Tilde{\bf p}} (\Tilde{\bf r},\Tilde{t}) \rangle$  are odd in ${\bf p}$ and thus vanish.  
The term $\langle  \partial_{\Tilde{\bf p}}  g_{\sigma\Tilde{\bf p}} (\Tilde{\bf r},\Tilde{t}) \rangle =0$  given that $g_{\sigma\Tilde{\bf p}} (\Tilde{\bf r},\Tilde{t})$ exponentially decays on the boundary of Brillouin zone.
Hence,  we get
\begin{align}
   & \partial_{\Tilde{t}} n_0 (\Tilde{\bf r},\Tilde{t})  \langle \bar{f}_{\Tilde{\bf p}}^0 \rangle
    + \langle \Tilde{\bf p} \cdot \partial_{\Tilde{\bf r}} g_{\sigma\Tilde{\bf p}} (\Tilde{\bf r},\Tilde{t}) \rangle  =0. 
\end{align}
Plugging in $g_{\sigma\Tilde{k}} (\Tilde{\bf r},\Tilde{t})$ and performing the integration we reach at
\begin{align}
   & \partial_{\Tilde{t}} n_0 (\Tilde{\bf r},\Tilde{t}) 
    =\frac{1}{1+\tau_C/\tau_d} \partial_{\Tilde{\bf r}} \cdot \left[\partial_{\Tilde{\bf r}} n_0 (\Tilde{\bf r},\Tilde{t}) + \partial_{\Tilde{\bf r}} \Tilde{V}(\Tilde{\bf r},\Tilde{t})  n_0 (\Tilde{\bf r},\Tilde{t})   \right] . 
\end{align}
Now we recall that $n_0 (\Tilde{\bf r},\Tilde{t}) =n (\Tilde{\bf r},\Tilde{t}) e^{\frac{\tau_0}{\tau_d}\Tilde{t}}$, and restore the original notations, resulting in Eq.~\eqref{eq:Drift-Diff} 
of the main text.

\vspace{10pt}

\end{widetext}
\end{document}